# Local properties of entanglement and application to collapse

Roland Omnès

Laboratoire de Physique Théorique (Unite Mixte de Recherche, CNRS, UMR 8627), Université de Paris XI, Bâtiment 210, 91405 Orsay Cedex, France

e-mail: roomnes@wanadoo.fr

Abstract: When a quantum system is macroscopic and becomes entangled with a microscopic one, this entanglement is not immediately total, but gradual and local. A study of this locality is the starting point of the present work and shows unexpected and detailed properties in the generation and propagation of entanglement between a measuring apparatus and a microscopic measured system. Of special importance is the propagation of entanglement in nonlinear waves with a finite velocity. When applied to the entanglement between a macroscopic system and its environment, this study yields also new results about the resulting disordered state. Finally, a mechanism of wave function collapse is proposed as an effect of perturbation in the growth of local entanglement between a measuring system and the measured one by waves of entanglement with the environment.

________

This is a revised version of arxiv.org.quant.phys.13093472, partly rewritten for clarity and with two previous key assumptions reconsidered, page 20: II as still convenient but unnecessary and III as being now proved.

It became clear early in the theory of quantum measurements –particularly in Schrödinger's works [1, 2]– that entanglement is the stumbling block forbidding emergence of a unique datum in a measurement. More recently many experiments confirmed this viewpoint by realizing conditions that were similar to a measurement but involved only a few atoms or photons [3]: Von Neumann's standard description of a measurement as a creation of entanglement was found perfectly adequate [4], even when there was decoherence, and there was as predicted no glimpse of collapse in these experiments.

An obvious consequence is that collapse, or the emergence of reality, is essentially a macroscopic phenomenon. This conclusion was drawn particularly in the theory of spontaneous collapse at a small but macroscopic scale by Ghirardi, Rimini and Weber [5], but in a form requiring modification in the quantum principles: The existence of a collapse effect with universal range and universal rate was added to these principles, the essential role of this effect being to break down the deadly obstruction from entanglement.

But one may also wonder whether entanglement is so total and rigid that a new principle is needed to break it. Is it impossible to look at it more ordinarily as a physical phenomenon needing time for its growth and space for its expansion, rather than remaining an absolute mathematical property of wave functions? This question can stand in some sense as the starting point of the present work.

When describing the nature of entanglement, Schrödinger considered an example where two quantum systems *A* and *B*, initially independent, begin to interact at some time zero and separate again after some more time [2]. Both systems are initially in a pure state but, although this is still true of the compound system *AB* after their interaction, it is not anymore true for each system separately. Schrödinger viewed this property of entanglement between wave functions as the most characteristic feature of quantum mechanics, the one estranging at the greatest degree quantum physics from classical physics. This strong standpoint led him to assert in a famous paper an unsurpassable opposition between the quantum principles and the uniqueness of measurement data [1].

One will try to remain close to Schrödinger's pattern in the present work by considering mostly a special example where a macroscopic system $B$, which is a Geiger detector containing an argon gas, interacts with an energetic alpha particle, denoted by $A$.

The first six sections in this paper deal with the case of a predictable measurement where the particle has a trajectory crossing the detector with certainty. The growth of entanglement is considered first in Section 2 as it can be envisioned in perturbation graphs for the atoms in the gas. The development of entanglement appears then gradual in this framework and can be seen as an increase in complexity for the topology of these graphs.

A convenient algorithm for Schrödinger's equation is built up in Section 3 to follow more explicitly this topological behavior of entanglement. This construction neither adds anything to the quantum principles nor subtracts anything from them, but whether it yields in principle the wave function of the $AB$ system at any time $t$, it keeps also memory of the past of entanglement between $A$ and $B$ until that time. In addition to its topological aspects, entanglement appears then as an evolving property in the history of wave functions and not so much of the wave functions themselves.

This property is expressed in terms of quantum field theory in Section 4 and the corresponding mathematical framework is sketched in Section 5. Little use will be made in practice of these sections in the present work and the main intent of Section 4 is only to indicate that the approach is general enough for the diversity of measurements and of measuring devices. The main point in Section 5, on the other hand, is to show that the local behavior of entanglement, although well defined through the history of the system, does not stand as a physical property of the wave function in Von Neumann's sense, which associates these properties with projection operators in Hilbert space [4].

As shown in Section 6, this peculiarity does not forbid the existence of a measure $f(x, t)$ expressing which proportion of atomic states in the detector became locally entangled with the alpha particle at time $t$ in the neighborhood of some space point $x$. One can also use kinetic theory to derive a transport equation for this measure of entanglement. This equation turns out to be nonlinear and shows a significant consequence, which is that local entanglement remains located behind the front of an entanglement wave progressing at a finite velocity (the sound velocity in a gas, Fermi velocity in a conductor or the velocity of light when entanglement is carried by photons).

These first six sections, which deal with a special case, make clear the meaning of local entanglement, which does not exhibit new physical effects but somewhat improves one's understanding of the marks in a macroscopic state of its past history. This is used in the last two sections to get more information about disorder in the state of a measuring device interacting with an environment, with emphasis on the fact that this interaction acted long ago before a measurement.

Section 7 turns attention to the interaction of the detector with its environment long before measurement. One deals again with a Geiger detector interacting with a standard external atmosphere. As well known, the main quantum effect is the occurrence of many entanglements of the gas in the detector with the atmosphere [6]. This is a very strong effect, as shown by decoherence when a measurement occurs. In Section 7, there is still no measurement but the discussion of partial entanglement in the previous sections is used to get a better understanding of disorder in the state of the detector with quantitative expression.

Such a state is there called "predecoherent", to stress that it built up before measurement whereas its origin and its strength are the same as for decoherence.

Section 8 deals at last with a quantum measurement by this detector in contact with the same environment. Two different kinds of local entanglement are then in presence: There is on one hand local entanglement of the detector with the alpha particle and there is predecoherence on the other hand, which results from an accumulation of past entanglements of the detector with the environment. The strength and behavior of predecoherence are known from Section 7 whereas the local growth of entanglement with the measured particle is known from Sections 3-6 One can study therefore how predecoherence perturbs the local growth of entanglement of the measuring apparatus with the measured particle.

The main point in that section and in the whole paper is finally a specific proposal for the origin of collapse: Intricacy (*i.e.*, local entanglement) is a property of the systems history, not of the quantum state at definite times. Its progress is irreversible. Intricacy between the measuring apparatus and the environment (*i.e.*, predecoherence) perturbs the progress of intricacy between the apparatus and the measured system. This perturbation has a strong influence on the entanglement of these two systems (if one means now entanglement in its usual sense as a property of the quantum state). The whole process brings out irreversible Brownian fluctuations in the quantum probabilities of measurement channels, ending inevitably with complete collapse. Quantitative evaluation of the corresponding time scale of collapse is very encouraging.

Section 9 contains conclusions together with some remarks concerning some relevant points in the interpretation of quantum mechanics.

## 2. Topological aspects of entanglement

One will deal mostly with the case of an alpha particle $A$ with a straight-line trajectory entering the Geiger counter $B$ at a sharp time 0. Figure 1 shows a perturbation graph for the events occurring in the detector before some time $t > 0$ (the same figure could represent equivalently a Feynman history during this same time interval, except that the straight lines representing propagation of the alpha particle and of argon atoms would become highly wiggling Feynman paths). The heavy horizontal line represents propagation of the alpha particle and the lighter horizontal lines represent propagation of atoms. Vertical lines connecting the particles at a definite time represent interactions.

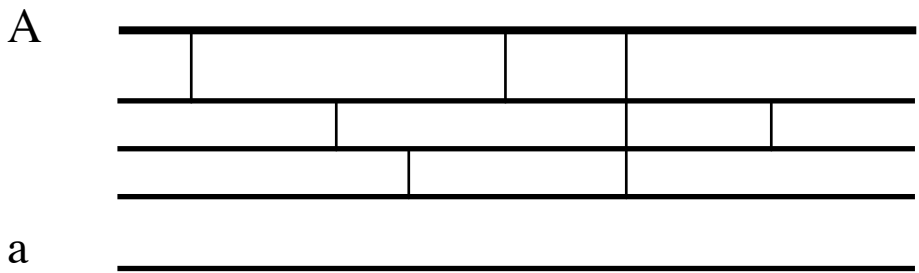

**Fig. 1** Topology of entanglement in a Feynman graph or a perturbation graph (time is going from left to right).

No explicit calculation is needed to see that, in this graph, some atoms have become entangled with the particle *A* at time *t* either because they interacted directly with *A* or interacted with an atom that had previously interacted with *A*, or interacted with an atom that had interacted with an atom that had interacted ... and so on. In Figure 1, an atom, denoted by *a*, is not connected with the alpha particle at time *t*, at least according to this graph.

This simple interpretation of a growth in entanglement has several significant consequences. First of all, it is clearly a topological property, either of perturbation graphs or Feynman histories. Secondly, it shows a strong analogy between the growth of entanglement and a contagion, since an atom can catch entanglement directly from the alpha particle but also catch it from an already entangled atom.

This topology of entanglement can be viewed also as a form of clustering if one considers that the set of entangled particles constitutes a cluster in the sense of graph theory. In that sense, the present trend of ideas is not new and occurred already in other domains of physics where it played a significant part. It first appeared in quantum statistical physics where it was used for instance to establish the proportionality of extensive thermodynamic quantities to the volume in a quantum framework [7, 8]. It occurred also in *S*-matrix theory [9] where it is often expressed as showing why an experiment in Geneva is insensitive to another experiment in Brookhaven. In the present case, it would mean that an atom far enough from the alpha track feels very little influence from this particle, at least during some time or far enough away. The formulation of these cluster properties was also thoroughly expressed in Weinberg's book [10], where it is was used among the foundations of effective quantum fields. Some other topological forms of clustering were also encountered in the scattering theory of several particles [11].

One could give a special name to these local and topological properties of entanglement and the name "intricacy" will be proper. Its topological character is not the only distinctive quality of intricacy. Feynman paths or perturbation graphs are not only drawings with a topology of connectedness but they carry mainly quantum amplitudes, which are quantitative attributes. These amplitudes add up and this addition will be studied in the next section dealing with wave functions.

The interest of intricacy for applications is not only concerned with wave functions however and its main consequences appear in a macroscopic system with disorder at a microscopic scale: the argon gas in the Geiger counter at finite temperature for instance. One will find in Section 6 that there exists in such a case a local measure of intricacy, expressing the probability for the gas atoms in any small space region to be intricate with the alpha particle at some time. Because of this locality of the measure of intricacy and to avoid playing with vocabulary, one will use most of the time the expression "local entanglement" rather than "intricacy" (except in Section 8 when the distinction between local and total entanglement will become central).

Another matter of language should also be mentioned to avoid misunderstanding: Although the connectedness of a specific atom in a perturbation graph is a sharp notion in that graph, where that atom is either intricate or not, the measure of local entanglement does not distinguish among atoms in the same region, but yields only a probability for these atoms to be connected with the alpha particle.

This is particularly clear for the first atoms to become entangled with the alpha particle: many of them become excited and a fraction of them is ionized. Even in the best textbooks describing this process, one speaks of "the" excited atoms or "the" ions and of their number. This is usually preferred to a more rigorous language in which one would speak of an "excited quantum state" or an "ionized state" for any undistinguishable atom or system of atoms, as well as the corresponding number operators in a small region and together with the corresponding probabilities. One will most often use the first language than the second one when dealing with intricacy and will speak of the number of intricate atoms or locally entangled atoms rather than speaking of the intricate state of any undistinguished atom in such or such region, together with the corresponding probability. Let one only hope that this common expression will create less confusion than there would tediousness in using a refined one.

**3. A Schrödinger equation for local entanglement**

Although the existence of local entanglement is most easily understood through perturbation theory, it can also be cast into the wider frame of Schrödinger equations. More precisely, one may devise a specific algorithm for solving the Schrödinger equation, through which one can get in principle not only the wave function itself at any time but also a memory of the growth of local entanglement up that time.

Let one fix first some notations: One denotes by $y$ the position of the alpha particle and a spin index if necessary [12], and let $N$ be the total number of argon atoms in the detector. These atoms are distinguished by an index $n$ ranging from 1 to $N$, the position of an atom being denoted by $x_n$ and the set of all these positions by $x$. ($x$ denotes therefore a point in configuration space). The initial state of the alpha particle $A$ is supposed a pure state with wave function $\chi(y)$ The initial state of the Geiger counter $B$, which is mixed, is described by a density matrix $\rho_B$. One concentrates first on one eigenvector of this matrix with wave function $\psi(x)$ so that the initial $A$-$B$ state is:

$$\Psi(x, y; 0) = \psi(x; 0)\, \chi(y; 0). \tag{3.1}$$

Its evolution is governed by the Schrödinger equation

$$i\hbar d\Psi/dt = (K_A + K_B + V_{A,B})\, \Psi, \tag{3.2}$$

where $K_A$ and $K_B$ denote respectively the kinetic energy of the alpha particle and of the argon atoms. One assumes for simplicity that all the interactions can be represented by potentials so that the potential energy $V_{A,B}$ is a sum $U + V$ where a potential $U$ represents the interactions between the alpha particle and the various atoms whereas $V$ is the sum of potential interactions between pairs of atoms:

$$U = \sum_n U(y, x_n), \qquad V = \sum_{nn'} V(x_n, x_{n'}). \tag{3.3}$$

In time-dependent perturbation theory, every perturbation term is associated with a graph as in Figure 1. In this graph, local entanglement is an irreversible topological property in so far as a definite atom is intricate once and for all with the alpha particle. Moreover, in this graph, any interaction at time $t$ of an already intricate atom with a not yet intricate one makes this second one intricate.

Since these topological properties of conservation and contagion of local entanglement are valid for every term in perturbation theory, it should be possible to find an algebraic formulation for them, To do so, one introduces an entanglement index for each atom with label $n$ (at this stage, the atoms are distinguished). This index takes the value 1 when the atom is intricate with the alpha particle) or the value 0 when the atom is not intricate. The generation of local entanglement of atoms by interaction with the alpha particle means that such an interaction provokes a transition 0 —> 1 in the index of a non-intricate atom whereas there is no such change and the transition is 1 —> 1 when the atom is already intricate. Similarly, an interaction between two atoms leads only to the transitions 00 —> 00, 01 —> 11, 10 —> 11 and 11 —> 11 between indices.

For a unique atom, the transitions 0 —> 0, 1 —> 1 and 0 —> 1 (there is no transition 1 —> 1) are respectively expressed by the $2 \times 2$ matrices

$$P_{n0} = (I + \sigma_z)/2, \ P_{n1} = (I - \sigma_z)/2, \ S = (\sigma_x + i\sigma_y)/2 \ , \tag{3.4}$$

which are written here in terms of Pauli matrices. $P_{n0}$ and $P_{n1}$ are projection operators, which conserve respectively a state of non-local entanglement or of local entanglement, whereas $S_n$ brings a state with no local entanglement to an intricate state.

Relying again on perturbation theory, one can express also algebraically the generation and the contagion of local entanglement through a rewriting of the potentials by $2 \times 2$ matrices for interactions with the alpha particle and $4 \times 4$ matrices for interaction between two atoms:

$$U(y, x_n) \longrightarrow U_n = U(y, x_n) A_n \ , \text{ with } \ A_n = (S_n P_{n0} + P_{n1}). \tag{3.5}$$

The first term $S_n P_{n0}$ in the $2 \times 2$ matrix $A_n$ describes the generation of local entanglement from a non-intricate atom $n$, this atom being recognized as non-entangled by the projection matrix $P_{n0}$ and brought to local entanglement by action of the matrix $S_n$. The term involving the projection matrix $P_{n1}$ in $A_n$ expresses that local entanglement is irreversible when an already intricate atom interacts (or interacts again) with the alpha particle.

The conservation or contagion of local entanglement in the interaction of two atoms $n$ and $n'$ is similarly expressed through a rewriting of the potential by $4 \times 4$ matrices, namely

$$V(x_n, x_{n'}) \longrightarrow V_{nn'} := V(x_n, x_{n'}) O_{nn'}, \tag{3.6}$$

with $\quad O_{nn'} = P_{n0} \otimes P_{n'0} + P_{n1} \otimes P_{n'1} + S_n P_{n0} \otimes P_{n'1} + P_{n1} \otimes S_{n'} P_{n'0} \tag{3.7}$

It becomes then clear that the evolution of entanglement is not restricted to perturbation theory but has a wider meaning. To formalize conveniently this extension, one can characterize a state of entanglement for the $N$ atoms by a string $q$ consisting of $N$ bits of entanglement indices taking the values 0 or 1. There are $2^N$ such strings. To each string, one associates a wave function $\Phi_q(\{x\}, y; t)$. The set of these wave functions can be considered as a $2^N$-dimensional vector $\Phi$ depending on $(\{x\}, y, t)$, with an evolution equation

$$i\hbar d\Phi/dt = H'\Phi \tag{3.8}$$

The operator $H'$ is not self-adjoint, because local growth of entanglement is not a reversible process. It is a $2^N \times 2^N$ matrix operator acting on the $2^N$ wave functions $\Phi_q$ according to

$$H'\Phi = (K_A + K_B) \Phi_q + \sum_{q'}(U_{qq'} + V_{qq'}) \Phi_{q'}, \tag{3.9}$$

The kinetic energy $K_A$ of the alpha particle is unchanged. The kinetic energy of atoms $K_B$ is also unchanged, except that it is now multiplied by the unit matrix $I_{qq'}$. The interaction $U_{qq'}$ between the alpha particle and atoms is again the sum of interactions $U_n$ as in (3.5), but multiplied by a $2^N \times 2^N$ matrix acting as the unit matrix on all the atoms $n' \neq n$ and as the matrix $A_n$ on atom $n$. Similarly, $V_{qq'}$ is a sum of potentials $V_{nn'}$ as in (3.5), each one of them being multiplied by a matrix acting as the identity on every atom $n''$, different from $n$ and $n'$, and as the matrix $O_{nn'}$ in its action on the indices of entanglement for the pair of atoms $nn'$.

The set of equations (3.8) can be used in principle in an algorithm for solving the Schrödinger equation (3.2). This procedure is much more involved than using an algorithm for solving directly (3.2, but it can yield much more information since it provides, at any time $t$, detailed information on the state of entanglement at that time. This information is moreover some sort of memory of the growth of entanglement as it occurred between the beginning of interaction of the alpha particle with the detector and time $t$.

Although the differential matrix operator $H'$ is not self-adjoint, this is because the generation and the contagion of entanglement are not time-reversible. Notwithstanding, the existence of solutions for (3.8) is essentially valid under the same very general conditions as the standard Schrödinger equation (3.2)if one assumes convenient bounds on the potentials and their derivatives [13]. Furthermore, a solution of (3.8) yields also the standard Schrödinger wave function $\Psi$ as the sum over the $2^N$ values of $q$ of the functions $\Phi_q$.

One can also account for the symmetry between indistinguishable atoms, like the Bose-Einstein symmetry of argon atoms in the present case. The Schrödinger wave function $\Psi$ is invariant under a permutation of atoms so that, in place of the $2^N$ functions $\Phi_q$, one can deal with a smaller set of $N+1$ symmetric functions $\Xi_r$ ($r = 0, 1, ..., N$) in which $r$ atoms are entangled with the alpha particle and $N - r$ atoms are not. The sum of these functions yields again the wave function $\Psi$.

**4. Mathematical aspects\***

The main point of this section is methodological and was briefly indicated in the Introduction. It can be therefore omitted in a first reading.

When local entanglement is introduced as a topological refinement in Schrödinger's equation, one may ask the meaning of this extension as far as the Hilbert space framework is concerned. One should first stress again that this idea of local entanglement (which is rather at the present stage a partial entanglement involving various numbers $r$ of entangled atoms) is closely linked with the macroscopic character of the measuring system $B$. One will find moreover in the next sections that the most interesting properties of partial entanglement are precisely its local properties, namely the distribution of partial entanglement in space and its evolution. In that sense, one is more concerned by wave functions than by abstract vectors in Hilbert space: If one could perform actual computations of these wave functions and look closely at their evolution, one would be probably able to notice many properties of interest with no trivial relation with the algebra of operators in Hilbert space. As a matter of fact, one will see in this section that local entanglement is one of these properties, with the special interest of being rather easily accessible to a non-Hilbertian analysis.

One will deal only with the case of symmetric functions $\Xi_r$. The evolution equation (3.8) is linear and one can therefore formulate it by using vector spaces. One may consider for instance that a wave function $\Xi_r$ is associated abstractly with a vector $| \Xi_r >$ in a linear space $E_r$ with definite symmetry properties for exchange between entangled and/or non-entangled atoms. Such a space $E_r$ inherits a scalar product from standard wave functions and a scalar product $< \Xi_r | \Xi'_r >$ of two functions $\Xi_r$ and $\Xi'_r$ in $E_r$ is well-defined as an integral of their product over the configuration space of atoms. A scalar product $< \Xi_r | \Xi'_{r'} >$ of two functions belonging to different vector spaces $E_r$ and $E_{r'}$ ($r \neq r'$) is also well defined. These two spaces are not orthogonal however, since such a scalar product does not generally vanish.

One may presume that the relevant mathematical framework is sheaf theory [14], the set $E'$ of $N$ vector spaces $\{E_r\}$ being a sheaf of Hilbert spaces. They communicate through an infinitesimal neighborhood of their common zero vector, a transfer of an infinitesimal vector from some $E_r$ to another $E_{r'}$ ($r' > r$) occurring during an infinitesimal time under (3.8). The topological aspects of local entanglement, which were encountered in perturbation theory, point also towards sheaf theory, which suits well also this kind of properties (as one can see from the well-known example of a sheaf of Riemann surfaces over a cut complex plane).

The standard Hilbert space $E$, to which standard wave functions $\Psi$ belong is the sum of vectors in the set $\{E.\}$, but there are significant differences: $E$ is the proper framework for expressing entanglement when there are several measurement channels. This entanglement, which one will call later "total entanglement" expresses then a vector in $E$ as a sum of tensor products between vectors in the Hilbert space of the alpha particle and vectors in the Hilbert space of the detector. This total entanglement, which is the only one to occur in standard measurement theory, extends from $E$ to $E'$, but local entanglement, which makes sense in $E'$, does not extend to $E$.

This mathematical, topological and physical difference between total –or standard– entanglement and local entanglement has a remarkable consequence, which is that local entanglement is not a standard physical property according to Von Neumann's mathematical definition of physical properties: No projection operator in Hilbert space can express local entanglement, no more that projections can account for the history of quantum evolution and only have only access to instantaneous properties of the state at a sharp time $t$ (except of course for conserved quantities).

This difference makes clearer the meaning of local entanglement from the standpoint of methodology: Its mathematical discussion, as was sketched here, does not reach any experimentally testable property of the system but more information about its past history, from the fact that this history was governed by the Schrödinger equation.

## 5. An approach using quantum field theory*

Several questions remained unanswered in the previous discussion. The main one is concerned with locality: If the alpha particle followed a straight-line trajectory in its initial state, its interactions with atoms will occur along this track and the contagion of entanglement will start from there. Since this contagion proceeds through collisions and the atoms have a finite mean free path, one expects this contagion to expand progressively farther away from the track. The algorithm in Section 3 does not yield these properties in an obvious manner however: It shows a gradual growth of entanglement but not obviously its locality. This is the main point of this section, which can be omitted in a first reading.

Several reasons suggest recourse to quantum field theory: One is looking for many-body properties, which are often best approached through a field version of quantum mechanics. The fact that an atom is brought from a state of no-entanglement to an entangled state through contagion suggests the combined action of an annihilation operator and a creation operator. Furthermore, a local field $\varphi(x, t)$ is particularly well able to exhibit locality.

Let one therefore briefly recall a few points in the field approach to the many-body problem [12]: The atoms are described by a field $\varphi(x)$, where the notation $x$ involves again the position of an atom and eventual spin indices. The field satisfies commutation or anti-commutation relations according to the spin value, but the two cases are very similar and one will retain only for illustration the case of Bose-Einstein statistics, which holds for argon atoms. The commutation relations are then

$$[\varphi(x), \varphi(x')] = 0, \quad [\varphi(x), \varphi\dagger(x')] = \delta(x - x'). \tag{5.1}$$

If one denotes the vacuum state by $|0>$, a state of a gas involving $N$ atoms with wave function $\psi(\{x\})$ is given by

$$|\psi> = \int\{dx\}\psi(\{x\})\prod_{r=1}^{N}\varphi\dagger(x_r)|0>, \tag{5.2}$$

(Notice the difference between the notation $x$ for localization of the field $\varphi(x)$ and the notation $\{x\}$ for all the variables in the wave function).

The field Hamiltonian is given by

$$H = \int dx\, \varphi\dagger(x)(-\Delta/2m)\varphi(x) + (1/2)\int dxdx'\, \varphi\dagger(x)\varphi\dagger(x')V(x, x')\varphi(x)\varphi(x') \tag{5.3}$$

where the factor 1/2 in the last term is due to the fact that a pair of atoms with positions $x$ and $x'$ occurs twice in this expression with the respective orderings $(x, x')$ and $(x', x)$).

To describe entanglement, one now introduces two fields $\varphi_0(x)$ and $\varphi_1(x)$, for non-entanglement and entanglement respectively. Both of them, together with their adjoint fields, satisfy the commutation relations (5.1). They are also supposed to commute together so that for instance,

$$[\varphi_0(x), \varphi_1(x')] = 0 \quad \text{and} \quad [\varphi_0(x), \varphi^\dagger_1(x')] = 0 \tag{5.4}$$

To get back the previous evolution of entanglement in wave functions, one must choose as before an operator $H'$ playing the part of an Hamiltonian in the evolution and yielding Equation (3.8) for the evolution of a wave function. This is obtained

$$H' = H_0 + H_1 + H_{01} + D_0 + D_1, \tag{5.5}$$

where $H_0$ and $H_1$ represent respectively the independent evolution of non-entangled and of entangled atoms; they have the same expression as (5.3) after replacing $)\varphi(x)$ by $\varphi_0(x)$ and $\varphi_1(x)$ respectively. The coupling $H_{01}$ representing the contagion of entanglement is given by

$$H_{01} = \int dxdx'\, \varphi^\dagger_1(x)\, \varphi^\dagger_1(x')V(x, x'\varphi_0(x)\,\varphi_1(x')\,. \tag{5.6}$$

The generation of entanglement by the alpha particle and its lack of change when the alpha particle interacts with an already entangled atom are described by two terms $D_0$ and $D_1$ in the Hamiltonian. They involve a field $\alpha(y)$ describing the alpha particle and are given by

$$D_0 = \int dxdy\, \alpha^\dagger(x)\,\varphi^\dagger_1(x)\,U(x, y)\alpha(y)\,\varphi_0(x), \tag{5.7}$$

$$D_0 = \int dxdy\, \alpha^\dagger(x)\,\varphi^\dagger_1(x)\,U(x, y)\alpha(y)\,\varphi_1(x). \tag{5.8}$$

There is however no direct relation between the basic field $\varphi(x)$ and the phenomenological fields $\varphi_0(x)$ and $\varphi_1(x)$. Nevertheless, one can show easily that Equation (3.8) for the evolution of locally entangled wave functions coincides with the effect of the field Hamiltonian (5.5), when Bose-Einstein symmetrized wave functions are used, as true in the present spinless case, and this coincidence confirms that the growth of local entanglement warrants its name and is definitely a local effect.

**6. Kinetic growth and transport of entanglement**

As a newt step, one can go from a quantum description of local entanglement to its macroscopic behavior. To that end and as often done in statistical physics, one covers the region of the gas inside the detector by a collection of small macroscopic cells (or Gibbs cells), denoted by $\beta$. Denoting by $N_\beta$ the number of atoms in such a cell, one can define a local measure of entanglement $f_{\beta 1}$ for the atoms in $\beta$, with

$$f_{\beta 1} = N_\beta^{-1} Tr\{\rho \int_\beta \varphi_1^\dagger(y) \varphi_1(y) \, dy\}. \tag{6.1}$$

A more cumbersome definition, which is nonetheless convenient for a better physical understanding, goes back to the local evolution of entanglement, as it was used in Section 3. One introduces the eigenfunctions $\Psi_n$ of $\rho$. Then, one writes down $\Psi_n$ as a sum of functions $\Phi_{nq}$ showing entanglement. From each one of these functions, one defines a probability of entanglement in $\beta$ at a time $t$ as the corresponding average value of the number of entangled atoms in $\beta$. Summing over $q$ and then over the eigenfunctions $\Psi_n$ and taking account of their probability in $\rho$, one obtains another explicit expression for $f_{\beta 1}$.

The number of eigenfunctions $\Psi_n$ is very large and the number of functions $\Phi_{nq}$ still much larger. Although the real parts of the scalar products $<\Phi_{nq} | \Phi_{n'q'}>$ do not vanish, they are so numerous and so erratic in sign that one can neglect their sum and consider accordingly this expression of $f_{\beta 1}$ as significant, and use it. In view of its construction and since the atoms are undistinguishable, one can consider $f_{\beta 1}$ as defining a probability for every atom in $\beta$ to be locally entangled. To make sense of this probability, one defines in the same way a probability $f_{\beta 0}$ for no entanglement through the relation

$$f_{\beta 1}(x, t) + f_{\beta 0}(x, t) = 1. \tag{6.2}$$

For simplicity, one considers the gas as in thermal equilibrium at some temperature $T$. The average velocity of atoms is then $v = (3k_B T/2m)^{1/2}$. One denotes by $\tau$ their mean free time $\tau$ and their mean free path by $\lambda$ equal to $v\tau$. At a macroscopic scale, local entanglement can be then considered as a kinetic effect: When an entangled atom collides with another atom, whther this second atom is entangled or not, the first atom conserves its entanglement though it suffers some change in its velocity from the collision. The second atom suffers also a change in its velocity, but becomes also entangled through the contagion of local entanglement.

Let one consider successively the dynamical effect of collisions and their contagion effect. It will be convenient for this purpose to stress the continuity of macroscopic quantities and use a notation $f_1(x, t)$ in place of $f_{\beta 1}(t)$, $x$ being for instance the center of the cell $\beta$. One defines similarly $\bar{f}_1(x, t)$ as $1 - f_1(x, t)$.

When an entangled atom collides with another atom, whatever the state of entanglement of this second atom, the change in motion of the first one generates a diffusion of entanglement, which can be described by the diffusion equation

$$(\partial f_1 / \partial t)_{diffusion} = D \Delta f_1 . \tag{6.4}$$

In later calculations, one will use the familiar expression from random walk, $D = (1/6)\lambda^2/\tau$, for the diffusion coefficient.

In addition to this transport of local entanglement, there is a local growth from contagion. The probability for a non-entangled atom in the neighborhood of $x$ to become entangled is during a short time interval $\delta t$ by contagion from an entangled atom is $f_1(x) \delta t/\tau$. The probability for an arbitrary to be non-entangled and become entangled by collision is therefore $f_1(x) \bar{f}_1(x)/\tau$. The local increase in entanglement is therefore given by

$$(\partial f_1 / \partial t)_{contagion} = f_1 \bar{f}_0 / \tau \tag{6.5}$$

Taking together the effects of diffusion and contagion and using (6.2), one obtains a nonlinear partial differential equation for the evolution of entanglement, which is:

$$\partial f_1 / \partial t = f_1(1 - f_1)/\tau + D \Delta f_1 . \tag{6.6}$$

In principle, one should add a source term in the right-hand side of the equation to account for generation of local entanglement through direct collisions with the alpha particle, but this extension can be considered as trivial.

Let one look then at the consequences of (6.6). There are restrictions on the function $f_1$, namely $1 \geq f_1(x, t) \geq 0$. They are strong constraints. In one dimension, one can show easily that they can be satisfied at most in a semi-infinite domain, but one skips the proof for brevity. Since the gas is in a finite region, $f_1$ can reach the bound 1 only in the restricted region where the alpha particle generates local entanglement and, there, the growth of entanglement stops and one is left with a standard diffusion equation, which conserves positivity. One must then concentrate on the condition $f_1(x, t) \geq 0$, or rather consider the surface $S$ on which $f_1$ vanishes. One is then left with equation (6.6) with the boundary condition $f_1 = 0$, $S$ being unknown.

To understand what happens, one considers a one-dimensional case, corresponding to a one-dimensional coordinate velocity of atoms $v' = 3^{-1/2}v$. On replaces the three-dimensional notation $x$ by a one-dimensional variable $z$. Let the source of entanglement stand far away on the left ($z = -\infty$), corresponding to the boundary condition $f_1(-\infty, t) = 1$. The physical nature of contagion in entanglement becomes then most helpful. Looking at the motion of atoms as a random walk with a mean free time $\tau$ between collisions and a length of free walk $3^{-1/2}\lambda$, one may consider on average over many collisions that two colliding atoms get away after collision with velocities $v'$ and $-v'$. This means that there exists a boundary of the walks spreading contagion, which moves at a velocity $v'$ towards increasing values of $x$.

Spontaneous generation of a wave front is a specific feature occurring only from nonlinear differential or partial differential equations [15], which is the case presently. Far from the source of entanglement $f_1(x, t)$ should move at velocity $v'$ and be a function $f_1(z, t) = g(z - v't)$. From (6.6), one finds a nonlinear differential equation for the function $g$. When taking the units of length and of time as the mean free path and mean free time, this equation is

$$3^{-1/2} dg/dz + g(1 - g) + (1/6) d^2 g/dz^2 = 0. \tag{6.7}$$

Figure 2 shows the solution of this equation when the front is at $z = 0$ (i.e., $g(0) = 0$) and the boundary condition $g(-\infty) = 1$ is used. The abscissa unit is a mean free path.

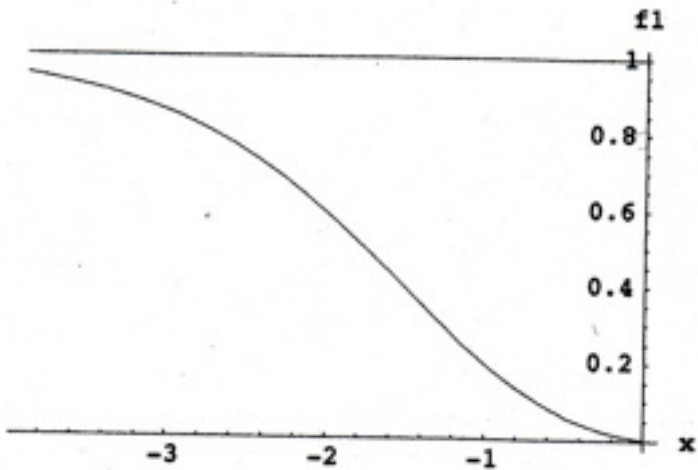

**Fig.2** An entanglement wave

This result is striking since it means that *entanglement grows locally behind a wave front moving at a finite velocity*. As a matter of fact, this velocity $v'$ coincides with the velocity of sound in a perfect gas and one may guess, by analogy between the carriers of entanglement in different situations, that entanglement waves would move at the Fermi velocity in a conductor or at the velocity of light when photons carry this entanglement.

One must take the result with care however. It would be too significant for not requiring much more careful and deeper investigations. Anyway, while acknowledging it as a conjecture, one will look now at the consequences it could have: perhaps nothing less than wave function collapse…

**7. The quantum state of an open macroscopic system**

*Disorder in an open system and predecoherence*

The interactions of an open macroscopic object with its environment are closely linked with their entanglement, as stressed particularly by H. D. Zeh and collaborators in the framework of decoherence theory [6]. The purpose of the present section is to elaborate on this relation and show that an account of the transient properties of local entanglement implies the existence of measure for the resulting disorder in the quantum state of the object, and to compute this measure.

One will use again as an example and a reference the case of the Geiger counter *B* containing an argon gas, when it is not performing a measurement or not yet performing one. The environment is ordinary atmosphere.

A few numbers will show better the situation. One assumes the gas and the atmosphere in thermal equilibrium at standard temperature and under the same standard pressure. The size *L* of the box containing the gas, its outer area *S* and its volume *V* are related by $V \approx L^3$, $S \approx L^2$ and *L* of the order of 10 *cm*. The number density of argon atoms in the box is of order $10^{19}$ per cubic centimeter and their total number of order $10^{22}$. The number of air molecules colliding on the box per second is then of order $10^{26}$. The front of an entanglement wave resulting from an individual collision has a velocity of order $10^5$ *cm/s* and it spends a time of order $10^{-4}$ second before filling up the whole gas and global entanglement is reached. The number of waves in the box at every time is therefore of order $10^{22}$, comparable to the number of atoms. The active front of a wave, in which local entanglement with the outgoing state is not complete and not vanishing is of the order of a few $10^{-5}$ *cm*, according to Figure 2. The number of these active regions at every point in the box is therefore of order $10^{18}$. One will say that such a situation, where the number of entanglement waves is comparable with the number of atoms, is a case of *complete disorder*, "complete" meaning then that nothing new would occur if the rate of interactions with the environment became still larger. A more precise definition will be given later.

Because the origin and strength of high disorder are close to those of decoherence, one will call its effect *predecoherence*.

*Predecoherence as a random process*

Every collision of an external molecule on the box occurs randomly in some small region of the surface at random times. One will consider these events as independent. Altogether, the collisions occurring during a time somewhat larger than the lifetime of the waves of local entanglements constitute a random process.

Random processes bring with them the notions of average and fluctuations and one begins by defining an average for the density matrix $\rho$ of the argon gas. Nothing in this state, certainly, is insensitive to the environment, except for its thermal equilibrium, and the average is necessarily

$$<\rho> = Z^{-1}\exp(-\beta H), \qquad (7.1)$$

where *H* is the Hamiltonian of the gas.

There are more complex situations where the object of interest has organization: in the case of a watch for instance, or when one accounts for the electric circuit in the Geiger detector. The notion of a quantity inaccessible to the effects of environment is then practically identical with the notion of accessible information and these quantities are slowly-varying average values $\{a_k\}$ of some set of macroscopic observables $\{A_k\}$, including usually positions and velocities. The average state (7.1) becomes then

$$<\rho> = \exp(-\beta H - \Sigma_k \lambda_k A_k) \qquad (7.2)$$

where the values of the Lagrange parameters $\lambda_k$ insure that this density matrix yields the average values $a_k$ for the observables $A_k$ and the identity matrix $I$ has been introduced among the observables $\{A_k\}$ to insure normalization of the trace through its average value [16].

In any case, the fluctuating part of the density matrix is then

$$\Delta\rho(t) = \rho(t) - <\rho>. \qquad (7.3)$$

The random matrix $\Delta\rho$ has a vanishing trace, since both $\rho$ and $<\rho>$ have a unit trace. It will be convenient to split it into two parts

$$\Delta\rho = \rho_+ - \rho_-, \qquad (7.4)$$

where $\rho_+$ and $-\rho_-$ involve respectively the positive and negative eigenvalues of $\Delta\rho$. The two matrices $\rho_+$ and $\rho_-$ are therefore positive and have the same trace, which one denotes by $K$.

The value of $K$ measures in some sense the degree of disorder in the quantum state of the system arising specifically from the local transport of entanglement with the environment. Whereas one can say that the density matrix $<\rho>$ represents already a complete disorder through its total lack of information, $\rho_+$ and $\rho_-$ show something different, which can be described from a historical standpoint as some memory of the recent history of interactions with the environment. As noticed previously, this memory is not an instantaneous Von Neumann property, expressible through a projection operator, but the quantity $K$ brings something completely new, which is the measure of a memory of entanglement, different from the previous local measures $f(x)$, which have become blurred up completely since there is a very high number of them and each one of them refers to entanglement from a unique collision.

This is remarkable: In principle, the existence of $\rho(t)$ has an actual meaning, since quantum mechanics would have no content concerning the object about which one is talking if one could not assert that this state exists. But to identify $\rho(t)$ with $<\rho>$ is to give up understanding of the action of environment, although one learned from decoherence that this action has much to do with the problems of measurement. One may thus expect that local properties of entanglement, or rather intricacy, plays a significant part in measurement theory, as will be shown in the next section.

Still more remarkable is the fact that the corresponding effect can be quantified, as expressed by the following

Proposition

When disorder in the state of a macroscopic system arising from the environment is complete, or rather maximal, one has

$$K = Tr(\rho_+) = Tr(\rho_+) = 4/3\pi. \qquad (7.5)$$

To be precise, this statement should rely on an explicit definition of complete disorder, which will be made clearer after some examination of the conditions leading to a proof of the proposition.

*Definition of complete disorder and proof of* (7.5)

One needs some tools and the main one is the notion of a Wigner random matrix, which one recalls [17]. A Wigner random matrix is an $N \times N$ self-adjoint matrix $W$ with $N$ large. The average values of its elements vanish. Different matrix elements are uncorrelated and the standard deviations $<|W_{jj'}|^2>$ of different matrix elements $W_{jj'}$ are equal. When $<|W_{jj'}|^2> = 1/N$, one will say that $W$ is the standard $N \times N$ Wigner matrix.

A basic theorem ("Wigner semicircle theorem") shows then that the probabilistic distribution of the eigenvalues $y$ of $W$ is

$$dP = (4 - y^2)^{1/2} dy. \qquad (7.6)$$

Local entanglement with the environment has certainly no effect on the distribution in energy of the system. One splits therefore the energy spectrum into a set of small intervals of width $\Delta E$, each one of them centered on some value $E$ of the energy. The probability for the total energy to be in that interval is therefore

$$\Delta P = Z^{-1} \exp(-\beta E) \Delta E, \qquad (7.7)$$

for both $\rho$ and $<\rho>$.

One denotes by $N$ the number of energy eigenvalues in the interval $\Delta E$. This number is very large but finite and, when $\rho$ and $<\rho>$ are restricted to this interval, they become $N \times N$ self-adjoint matrices $\rho_N$ and $<\rho_N>$. It will be convenient to renormalize temporarily their trace to unity The eigenvectors of $<\rho_N>$ coincide with eigenvectors $|j>$ of the Hamiltonian and one has, in this basis,

$$<\rho_N> = (1/N) I_N, \qquad (7.8)$$

where $I_N$ is the unit $N \times N$ matrix.

Let one denote by $|n>$ the eigenvectors of $\rho_N$ and by $p_n$ the corresponding eigenvalues. One defines then explicitly complete disorder by the two following conditions;

1. The eigenvalues of $\rho_N$ are random and the random positive numbers $p_n$ have a Poisson distribution with average value $1/N$.

2. The unitary matrix bringing the eigenvectors $|n>$ of $\rho_N$ onto the basis $\{|j>\}$ is random.

The average value $1/N$ for the quantities $p_n$ insure a correct value for the trace of $\rho_N$. The Poisson distribution is also assumed because it minimizes the content of information for a distribution of positive quantities. As for Assumption 2, it means that the average value of a scalar product $<n|j>$ vanishes and, furthermore, if one denotes by $E(a)$ the average value of a quantity $a$ under the random orientation of the eigenvectors of $\rho_N$, one has

$$E(<j|n><j'|n'>^*) = \delta_{jj'} \delta_{nn'} / N. \qquad (7.9)$$

When averaged upon the Poisson distribution of its eigenvalues in its own basis of eigenvectors, $\rho_N$ yields an average $(1/N) I_N$, but since the identity matrix has the same form in every basis, the average of $\rho_N - <\rho_N>$ vanishes and $\Delta\rho_N$ is a pure fluctuation.

Introducing the fluctuations $\delta p_n = p_n - 1/N$ and denoting by $\Delta\rho_{Njj'}$ a matrix element $<j|\Delta\rho_N|j'>$, one gets:

$$\Delta\rho_{Njj'} \Delta\rho_{Nkk'} = \Sigma_{nn'} <j|n>\delta p_n <n|j'><k|n'>\delta p_{n'} <n'|k'>. \quad (7.10)$$

But according to (7.9), an averaging over orientation of the vectors $|n>$ yields only non-zero values when $j = k$, $j' = k'$ and $n = n'$. The average over the Poisson distribution of eigenvalues yields on the other hand, when taking account that the Poisson averages $<(\delta p_n)^2> = <p_n> = 1/N$:

$$E(\Delta\rho_{Njj'} \Delta\rho_{Nkk'}) = \delta_{jj'}\delta_{kk'}/N. \quad (7.11)$$

The matrix $\Delta\rho_N$ is therefore a standard $N \times N$ Wigner matrix and the distribution of its eigenvalues is given by (7.6). The average of its positive eigenvalues is then $(4/3\pi)/N$ and the opposite for the average of the negative eigenvalues, from which one gets

$$Tr(\rho_{N+}) = Tr(\rho_{N-}) = 4/3\pi. \quad (7.12)$$

Dropping the temporary renormalization and summing over the intervals $\Delta E$, one obtains finally the announced result (7.5).

To conclude, one acknowledges that, when defining complete disorder through the orientation of eigenvectors and the distribution of eigenvalues, the necessary conditions are not easy to validate explicitly. There is probably a tendency of the traces (7.12) to increase from small values when the interaction with the environment is weak and tend asymptotically to the value (7.12) when interactions become strong. To this, one might add considerations on enlarging the environment to an environment of the environment and/or using arguments, inspired by Von Neumann's chain of apparatuses measuring apparatuses [4]. But one will avoid that. One will simply consider the asymptotic value (7.12) as a sensible approximation in the case of the detector one is discussing.

(*Note*: This approach bypasses a less informative previous introduction of $\rho_+$ and $\rho_-$, which was moreover partly wrong after an erroneous introduction of some arbitrary phases in Equation (3.12) of that work [18]).

## 8. Collapse of quantum states

*Framework*

One turns now to quantum measurements and more precisely to real quantum measurements. A real measurement differs from the abstract description in most textbooks by the fact that it involves *three* interacting physical systems: There is a measured system *A*, usually microscopic, which carries a quantum observable *Z* that must be measured. There is also a measuring apparatus *B*, which will be considered as necessarily macroscopic. The interaction between *A* and *B* is special and such that the apparatus evolution is extremely sensitive to the value of *Z*, so much that it behaves at large scale in very different ways for different values of *Z*. Last but not least, there is an environment or, more properly, the outside universe in which a unique macroscopic reality is present.

For definiteness, one will deal again with an example where the measuring device *B* is a Geiger detector containing an argon gas. The measured system A is an energetic alpha particle, initially in a pure superposed state

$$|\psi> = c_1|1> + c_2|2>. \tag{8.1}$$

Both states $|1>$ and $|2>$ represent an alpha particle following a straight-line trajectory, which crosses the detector in state $|1>$ and does not cross it in state $|2>$. One will say that Channel 2, where nothing happens, is a "mute" channel. Two channel probabilities are the defined as $p_1 = |c_1|^2$, $p_2 = |c_2|^2$.

There are several motivations for choosing an example where one channel is mute, first because it makes the discussion shorter and some mathematics are less cumbersome. In addition, it keeps a door open for a later account of non-separability in quantum mechanics, which would enter inevitably the discussion if another detector were ready to detect the particle along its other trajectory in state $|2>$.

The case of an isolated apparatus is well known and one needs not discuss it, except for noticing that an initial generation of excited atoms, ions and electrons along the particle track in channel 1 is a starting point for intricacy and its contagion.

*A proposal*

A unique reality is not the only relevant character of the environment, when a quantum measurement occurs. Another character, almost opposite, is also the strong quantum disorder holding almost everywhere in it. One saw in the last section how local entanglement allows some understanding of this disorder. One knows also since Schrödinger's enquiries that complete entanglement is the main obstacle against understanding why a unique datum comes out in a measurement [1].

A clean distinction between these two aspects of entanglement is therefore convenient and, from there on, one will restrict the name "entanglement" to its standard meaning and speak of "intricacy" where the expression "local entanglement" was used previously.

To be more precise, entanglement expresses essentially that every eigenvector of the density matrix $\rho_{AB}$ can be written as a superposition

$$|\psi>_{AB} = |\chi_1>_B \otimes |1>_A + |\chi_2>_B \otimes |2>_A. \tag{8.2}$$

This is a mathematical concept relying on the property that the two states $|1>$ and $|2>$ of the measured particle occur in the present case (as a matter of fact, the wave function of the alpha particle evolves in the detector under slowing down, but one can easily account for that by making wider the meaning of the state vector $|1>_A$, which is no more unique).

On the other hand, intricacy was already discussed in some detail in the present work, most often under the name of "local entanglement" and one saw its multiple aspects: topological, historical as a memory of past interactions and of course evolving, local and measurable. Another property, obvious in its formulation, is also that intricacy is irreversible. This aspect appeared already when Schrödinger introduced entanglement as a property of two systems *after* they interacted [2]. One encountered it also in Section 3 when noticing that intricacy evolves, according to (3.8), under the action of a non self-adjoint operator *H'*. Finally, one found in Section 6 that intricacy spends a finite time to cross the detector, after which it leaves the whole place to entanglement. Like intricacy is a memory of past interactions, a final state of entanglement is also from the same standpoint a frozen memory that intricacy came to completion.

This distinction between entanglement and intricacy allows a clean statement of the proposal of the present work, which is:

MAIN PROPOSAL

*Collapse is a quantum phenomenon, originating in fluctuations of channels probabilities. These fluctuations are generated by an action of predecoherence on the growth of intricacy between the measured system and the measuring apparatus, predecoherence being due by itself to an accumulation of previous intricacies of the apparatus with the environment.*

The relation between intricacy and entanglement is especially clear in the case of the alpha particle in the state (8.1). One can then define more directly a state of the apparatus that is entangled with state $|1>$ as belonging to the matrix $<1|\rho_{AB}|1>$. It is clear, then, that the state of an atom can only be intricate with $|1>$ if it is entangled with it. On the other hand, state that is entangled with $|2>$ belongs to $<2|\rho_{AB}|2>$, but no state is intricate with $|2>$ since there is no *AB* interaction in that state.

In practice, one will show that that the main effect of the fluctuations in the growth of *AB*-intricacy, which are mentioned in the main proposal, yield fluctuations in the channel probabilities when they are combined with entanglement.

*Description of predecoherence in a measuring apparatus*

Predecoherence was described in Section 7 as a measurable property of disorder in the state of the apparatus before measurement. One found then that there exists two matrices, $\rho_+$ and $\rho_-$, which express qualitatively the existence of a disorder that is not purely thermal and, quantitatively, the strength with which the environment affects the quantum state of the apparatus. These matrices have therefore a significant physical meaning and one would like to find which part they could play in collapse.

To begin with, one must extend the results of Section 7 to the new situation occurring when some interaction of the apparatus with the alpha particle has begun at some time, which one takes for time zero. This interaction occurs of course with state $|1>$. At a later time *t*, one will write down the density matrix $\rho_{AB}$ describing the compound system *AB* as was done for the density matrix $\rho$ in (7.2-3), which one denotes now by $\rho_B$.

The expression of the density matrix showing the effect of predecoherence is then:

$$\rho_{AB}(t) = <\rho>_{AB}(t) + \rho_{AB+}(t) - \rho_{AB-}(t). \tag{8.3}$$

But the matrix $<\rho>_{AB}$ cannot be defined as an average over the random process of predecoherence as it was previously, since the main proposal implies that the randomness of collapse should originate in this predecoherent randomness. One cannot therefore use this kind of average when considering the possibility of a unique datum in a unique measurement. So, how can one define $<\rho>_{AB}$?

If a theory can confirm the main proposal, it must master quantitatively the amount of predecoherent disorder and rely on an explicit theoretical expression for this disorder. This expression is given in (8.3) the difference $\rho_{AB+} - \rho_{AB-}$, in such a way that no predecoherent disorder is present in $<\rho>_{AB}$. Furthermore, the proposal also relies on intricacy between the apparatus and the measured system, so that one should consider $\rho_{AB}$ in the framework of intricacy.

Should one then go as far as working with a sheaf of non-mutually orthogonal Hilbert spaces as one did in Section 5? If that were necessary, one could as well give up the hope of making the proposal into a theory. The only hopeful approach consists thus in considering that $<\rho>_{AB}$ involves no part of the disorder in predecoherence. It should carry however the channel probabilities $(p_1, p_2)$, even if they differ from their initial values $=|c_1|^2$, $p_2 = |c_2|^2$. Finally, is also associated with a definite amount and repartition of $A$-$B$ intricacy, since these are kinetic quantities, independent of any history.

One therefore defines $<\rho>_{AB}$ through the expression:

$$<\rho>_{AB}(t) = p_1(t) <\rho>_{B1}(t) \otimes |1><1| + p_2(t) <\rho>_{B2}(t) \otimes |2><2|, \qquad (8.4)$$

The matrix $<\rho>_{B1}$ does not depend on predecoherence and therefore not on the environment. It started at time zero from $<\rho>_B$ in (7.1), had no interaction with the environment and interacted only at times $t > 0$ with the alpha particle in state $|1>$. This interaction generated first a track of excited atoms, ions, electrons, and thereby a contagion of intricacy. One assumes that at least the kinetic amount of this intricacy is associated with $<\rho>_{B1}$ as a memory of its history.

The matrix $<\rho>_{B2}$, which corresponds to the mute state $|2>$ of the alpha particle, is very simple, since it is a density matrix of thermal equilibrium with no intricacy in its past.

One notices also that there should be no non-diagonal parts, such as $<1|<\rho>_{AB}|2>$, in (8.4), because $<1|<\rho_{AB}|2>$ is very sensitive to the environment, as known from decoherence theory.

As for the coefficients $p_1(t)$ and $p_2(t)$ in (8.4), they can be considered as objective quantities, namely the traces of $<1|\rho_{AB}|1>$ and $<2|\rho_{AB}|2>$. Their values in (8.4) are therefore borrowed from the actual density matrix $\rho_{AB}(t)$, in which they are supposed known.

As in Section 7, one will introduce the matrices

$$\Delta\rho_{AB}(t) = \rho_{AB}(t) - <\rho>_{AB}(t), \qquad (8.5)$$

with $\quad \Delta\rho_{AB}(t) = \rho_{AB+}(t) - \rho_{AB-}(t), \qquad (8.6)$

the matrices $\rho_{AB+}$ and $-\rho_{AB-}(t)$ being respectively the parts with positive and negative eigenvalues of $\Delta\rho_{AB}$.

*Three assumptions*

A specific kind of quantum disorder, originating in the environment, belongs to predecoherence. It is completely expressed by the two matrices $\rho_{AB+}$ and $\rho_{AB-}$, but one needs to know or to guess more about them to understand which action they involve. To do so, one points out to start with three simple properties of these matrices, which are expected valid to some degree and could lead directly to significant consequences. When stated as Is, these properties are as follow:

Assumption I:

One considers a measuring system in which predecoherence from the environment is strong and corresponds approximately to the trace properties

$$K = Tr(\rho_{AB+}) = Tr(\rho_{AB-}) = 4/3\pi. \tag{8.7}$$

Assumption II:

The diagonal sub-matrices of $\rho_{AB+}$ and $\rho_{AB-}$ according to the measured states $|1\rangle$ and $|2\rangle$ satisfy the trace properties

$$Tr<1|\rho_{AB+}|1> = Tr(<1|\rho_{AB-}|1>) = (4/3\pi)p_1, \tag{8.8a}$$

$$Tr<2|\rho_{AB+}|2> = Tr(<2|\rho_{AB-}|2>) = (4/3\pi)p_2. \tag{8.8b}$$

Assumption III:

This assumption is concerned with intricacy and consists of two parts:

III*A*: Local measures of intricacy with state $|1\rangle$, denoted by $f_{1+}(x)$ and $f_{1-}(x)$, make sense for the matrices $\rho_{AB+}$ and $\rho_{AB-}$. Their values are equal and coincide with the value $f_1(x)$ that they have in both $\rho_{AB}$ and $<\rho>_{AB}$.

III*B*: Because of the algebraic complexity of extracting eigenvalues and eigenvectors from $\Delta\rho_{AB}$, no criterion can distinguish whether a non-intricate state in $\rho_{AB+}$ (or $\rho_{AB-}$) is entangled with $|1\rangle$ or with $|2\rangle$.

Let one then comment these assumptions: Assumption I is not essential and, as one will see when coming to applications, the only significant property of the traces (8.9) is that they should be of order $O(1)$ or at least not very much smaller.

Assumption II is more a convenience than a condition. It came from a study of the algebraic process of extracting positive and negative eigenvalues of the difference $\Delta\rho_{AB}$, but with no satisfactory proof. Fortunately, this assumption is unnecessary and the collapse effect that will be described does not depend on Assumption II. One needs only to know, for instance, that $Tr<1|\rho_{AB+}|1>$ is equal to $Tr(\rho_{AB+})$ when $p_1 = 1$ and vanishes with $p_1$. Otherwise, $Tr<1|\rho_{AB+}|1>$ can vary with time more or less arbitrarily and even randomly without changing the forthcoming predictions regarding collapse. Equations (8.10) represent in that sense a convenient interpolation between the extreme values 1 and 0 for $p_1$ (and similarly for $p_2$). They make the forthcoming calculations slightly simpler but the main reason for using them here is that the present remarks were only made when this paper was sent to printing.

Assumption III looked at first as most difficult to prove or even to investigate, in so far as intricacy is not a Von Neumann property holding in Hilbert space. There is however a remarkably simple way to justify it, relying on the existence of progressive intricacy waves and anticipating on the very short time scale of collapse:

Let one assume first that interaction of the apparatus with the environment stops completely at times $t > 0$ when the measuring $AB$ interaction begins acting. The evolution of $\rho_{AB}$ is then unitary and governed by the $AB$ Hamiltonian. The same is true for $<\rho>_{AB}$ and for the eigenfunctions of $\rho_{AB+}$ and $\rho_{AB-}$. But one saw earlier that the growth of intricacy is essentially a kinetic property, so that its macroscopic measure $f_1(x)$ is practically independent of the density matrix on which it is built. If this independence holds for $\rho_{AB+}$ and $\rho_{AB-}$, Assumption IIIA can be considered as valid when interactions with the environment have been switched off at time 0.

But one saw also that predecoherence is generated continuously by external collisions, of which the effects are transported with a finite velocity along intricacy waves. Far enough from the external box, inside the apparatus, $\rho_{AB+}$ and $\rho_{AB-}$ are no yet affected by external collisions, which arrived on the box after time zero. This immunity lasts for some time at every place, depending on the distance to the boundary but, if collapse has a very short time scale, it will have occurred in the depth of the apparatus before arrival of new external perturbations. This argument, which bypasses in such a simple physical way especially tricky mathematical problems, is intriguing, perhaps amusing, but certainly inspiring. One will not judge it but look anyway at its consequences.

As for IIIB, which is not an assumption but a statement, it expresses the fact that no projection operator in Hilbert space can identify a quantum state as being intricate or not. It is essential in the present theory of collapse.

Finally, one should mention that in the case of a mute channel 2, there is no $AB$ interaction, no intricacy and no measure of intricacy so that $f_2(x)$ is identical to zero everywhere.

*Mechanism of probability fluctuations*

One now considers how the proposed mechanism for fluctuations in the channel probabilities would work.

Since intricacy is a local effect, one concentrates on a small macroscopic cell $\beta$, centered at a point $x$. In place of the notation $f_1(x)$ for the local measure of intricacy, one uses its average $f_{\beta 1}$ in $\beta$. If $N_\beta$ denotes the average number of atoms in the cell, the average number of intricate atoms in $\beta$ is therefore $f_{\beta 1} N_\beta$.

Since intricacy grows through generation and contagion, one considers first its generation from interactions of the alpha particle with atoms in the gas, necessarily in Channel 1. These interactions is well known from the Bethe-Heitler theory: Excited atoms are produced along a track in a neighborhood of the particle trajectory, together with fewer ions and free electrons. Some atoms suffer also scattering from the moving electric field of the particle through Van der Waals forces, but the corresponding probability is much smaller than the probability of excitation. Some time later, free electrons are accelerated by a static electric field in the detector and produce a cascade of secondary ions and electrons. Some photons are also emitted from decay of excited atoms. These various effects will be mentioned again when quantitative estimates will be made, but one disregards them presently to make the discussion clearer and one concentrates attention on excited atoms, which have the highest initial quantum probabilities and constitute the first generation of intricate atoms. Their average number in a cell is small and it even vanishes when the cell is far enough from the track. It is given initially by the product $f_{\beta1}(0)N_\beta$, with $f_{\beta1}(0)$ everywhere small or zero in the various cells $\beta$.

To study the evolution of intricacy and the associated evolution of entanglement, one will look at their behavior in the different matrices $<\rho>_{AB}$, $\rho_{AB+}$ and $\rho_{AB-}$ in (8.3).

The matrix $<\rho>_{AB}(t)$ shows up the values of $(p_1, p_2)$ according to (8.4) It also shows up intricacy, as explained in its construction. The measure of intricacy in $<\rho>_{B1}(t)$ is $f_{\beta1}(t)$ and it vanishes in $<\rho>_{B2}(t)$. Again by construction, the evolution of these matrices is insensitive to environment, so that intricacy in $<\rho>_{B1}$ grows by contagion as it did in Section 6 and its change, during a very short time interval $[t, t + \delta t]$, is

$$\delta f_{\beta1} = f_{\beta1}(1 - f_{\beta1})(\delta t/\tau), \tag{8.9}$$

where $\tau$ is the mean free time of atoms. Intricacy in $<\rho>_{B2}$ remains on the other hand zero.

The action of predecoherence from the environment is then contained entirely in the two matrices $\rho_{AB+}$ and $\rho_{AB-}$. In order to account for local properties, one introduces a density matrix $\rho_\beta$ for the cell $\beta$ by tracing out from $\rho_{AB}$ everything outside $\beta$. The same operation defines then two localized matrices $(\rho_{\beta+}, \rho_{\beta-})$.

Let one consider first $\rho_{\beta+}$ and take its trace $K$ as $4/3\pi$. The trace $T_{\beta2+}$ of $<1|\rho_{\beta+}|1>$ is $Kp_1$ and the trace $T_{\beta2+}$ of $<2|\rho_{\beta+}|2>$ is $Kp_2$, according to Assumption II. According to Assumption IIIA, $\rho_{\beta+}$ involves also a measure of intricacy $f_{\beta1}$ and carries a probability of intricacy $K p_1 f_{\beta1}$. According to Assumption IIIB, there is also a probability $K(1 - p_1 f_{\beta1})$ for non-intricacy, with no distinction between the two channels.

Because of the absence of any difference between the two channels as far as non-intricacy is concerned, intricacy grows when an intricate atomic state (or an intricate atom $a$ to say it shortly) interacts with a non-intricate atom $a'$, whether the state of $a'$ is entangled with $|1>$ or with $|2>$. Whereas the probability for intricate atomic states $a$ is $K p_1 f_{\beta1}$, the probability for non-intricate states $a'$ is $K(1 - p_1 f_{\beta1})$, which can be written also as

$$p_{\beta0} = K [p_1(1 - f_{\beta1}) + p_2]. \tag{8.10}$$

According to Assumption IIIB, (8.10) means that there is a probability $Kp_1(1 - f_{\beta1})$ for a non-intricate state $a'$ to belong to the matrix $<1|\rho_{\beta+}|1>$ and a probability $Kp_2$ for it to belong to $<2|\rho_{\beta+}|2>$. A collision between an intricate atomic state *a* (which belongs necessarily to $<1|\rho_{\beta+}|1>$) with a non-intricate atomic state *a'* belonging to $<2|\rho_{\beta+}|2>$ makes this state *a'* intricate and therefore belonging to $<1|\rho_{\beta+}|1>$. *This transition, which is due to the existence of intricacy, fractures the barrier between channels, which was established by entanglement. It is the source of probability fluctuations in the present approach, and hence of collapse.*

The increase in intricacy following from this effect arises from individual collisions, which on assumes incoherent in $\rho_{\beta+}$ in view of the dominance of predecoherent disorder in this matrix. There is therefore an increase in the trace $T_{\beta 1+}$ of $<1|\rho_{\beta+}|1>$, which is

$$\delta T_{\beta 1+} = K f_{\beta 1} p_1 p_2 \, \delta t / \tau. \tag{8.11}$$

The same effect occurs in $-\rho_{\beta-}$, but with an opposite sign (this difference in sign does not mean an intervention of negative probabilities, but only infinitesimal negative increases in the trace $T_{\beta 1}$ of $<1|\rho_\beta|1>$ during an infinitesimal time $\delta t$). The effects of $\rho_{\beta+}$ and of $\rho_{\beta-}$ cancel each other on average.

Incomplete cancellation between these effects of $\rho_{\beta+}$ and $\rho_{\beta-}$ can yield however fluctuations and this question reaches now the foreground.

In view of incoherence, individual atomic collisions, during the time interval $\delta t$ and according to the matrices $\rho_{\beta+}$ and $\rho_{\beta-}$, are independent events. The corresponding probabilities have therefore uncorrelated Poisson distributions. The difference $\delta T_{\beta 1} = \delta T_{\beta 1+} - \delta T_{\beta 1-}$ is then a random quantity with average value zero. Standard deviations and correlations between the two channels are given by

$$<(\delta T_{\beta 1})^2> = <(\delta T_{\beta 2})^2> = - <\delta T_{\beta 1} \delta T_{\beta 2}> = (8/3\pi) f_{\beta 1} p_1 p_2 \delta t / \tau. \tag{8.12}$$

A factor 2 in the right-hand side is due to the independent contributions of $\delta\rho_{\beta+}$ and $\delta\rho_{\beta+}$. One also used the relation $\delta T_{\beta 1} = -\delta T_{\beta 2}$ resulting from conservation of the unit trace of $\rho_\beta$. (When introducing the factor $p_1 p_2$ in the right-hand side, one used Assumption II, but the meaning of this factor as representing only an interpolation should be clear.)

Local fluctuations in different cells $\beta$ imply global fluctuations in the channel probabilities $(p_1, p_2)$. To compute these significant quantities, one writes down the full density matrix $\rho_{AB}$ as a product:

$$\rho_{AB} \approx \prod_\beta^\otimes \rho_\beta \quad \rho_{AB} \approx \Pi_\beta \rho_\beta. \tag{8.13}$$

This expression could appear of course a very rough approximation, but one will use it only for fluctuations and will return to its meaning afterwards.

Considering the probability $p_1$ as the trace of $<1|\rho_{AB}|1>$ and remembering that $\delta T_{\beta 1}$ is the trace of $<1|\delta\rho_{\beta+} - \delta\rho_{\beta-}|1>$, one gets standard deviations and correlation for the fluctuations in probabilities:

$$< (\delta p_1)^2 > \ = \ < (\delta p_2)^2 > = - < \delta p_1 \delta p_2 > \ = \ (8/3\pi)\, p_1 p_2 (\delta t/\tau) \sum_\beta f_{\beta 1} \ . \qquad (8.14)$$

This is the main result of the present work, to which one will only add two remarks.

The first one is concerned with a justification of using the factorization (8.13) when computing fluctuations. It consists in estimating errors in local fluctuations: A cell $\beta$ is a region of space and the main relative errors entering in its properties arise from fluctuations in $N_\beta$. They are of order $N_\beta^{-1/2}$. The relative error in (8.14) can be then estimated as of order $(N'N_\beta)^{-1/2}$, were $N'$ is the number of cells in which intricacy does not vanish. The error is negligible. On the other hand, correlations between neighboring cells bring also random errors in summations, but they will be also considered as negligible.

The second remark consists in generalizing (8.14) to any number of channels. This is easy and the key point is only to notice that, when there are several active channels $j$, no-intricacy means no intricacy with any channel. One gets then generally:

$$< \delta p_j \delta p_{j'} > = - (8/3\pi)\, p_j p_{j'} (\delta t/\tau)/\tau) \sum_\beta (f_{\beta j} + f_{\beta j'})(1 - \sum_{k \neq j, j'} p_k f_{\beta k}) \ \text{for}\ j \neq j', \qquad (8.15)$$

where the factor $1 - \sum_{k \neq j, j'} p_k f_{\beta k}$ stands for the total probability for no intricacy with the two channels $(j, j')$. The standard deviations are given then by

$$< (\delta p_j)^2 > = - \sum_{j' \neq j} < \delta p_j \delta p_{j'} > \qquad (8.16)$$

These results are valid also when there are mute channels (denoted by indices $l$), for which one must write $f_{\beta l} = 0$.

*Quantitative estimates*

One will consider explicitly a unique quantitative example. It deals with the detector for which some estimates were already made at the beginning of Section 7. The measured system consists in an alpha particle with energy 10 *MeV*, which crosses the detector in Channel 1. It leaves then a track with a length of order 10 *cm* before being slowed down.

The average excitation energy of excited atoms along the track is about 10 *eV* for argon. The total number of these atoms is therefore about $10^6$ and their average separation $l$ is therefore $10^{-5}$ *cm*. The mean free path $\lambda$ of an argon atom is also of order $10^{-5}$ *cm* and the mean free time of order $10^{-10}$ *s*. The size $\Lambda$ of the $\beta$ cells must be significantly larger than $\lambda$ for the factorization in (8.13) to make some sense. Applying Equation (8.14), one gets

$$< (\delta p_1)^2 > = < (\delta p_2)^2 > = - < \delta p_1\ \delta p_2 > = A\, p_1 p_2 \delta t \ , \qquad (8.17)$$

where the coefficient $A$ is of order $10^{11}\ (l/\Lambda)^2\ s^{-1}$.

In view of the rather small ratio $l/\Lambda$ this rate of fluctuations in the channels probabilities may look rather small for leading soon enough to complete collapse. But the situation evolves drastically soon after, particularly in a realistic detector where an electric field produces a cascade of secondary ionizations and excitations. Every measure of intricacy $f_\beta$ increases exponentially with time along the track, and intricacy extends all of them farther away from the track during contagion. The scene of action where probability fluctuations are produced and the rate of their growth grow tremendously. The scene is essentially an inflating cylinder with axis on the track. When the radius of this cylinder reaches the size $\Lambda$ of macroscopic cells, the measure $f_\beta$, which started from a value of order $10^{-6}$ (if $\Lambda = 10\lambda$) becomes close to 1 and the coefficient *A* in (8.17) reaches then a value of order $10^{16}$ $s^{-1}$. This highly macroscopic situation is reached about $10^{-9}$ *s* after the particle entry into the detector.

Other situations were considered but they will not be reviewed here. Their results seem in agreement with the known requirements on a spontaneous collapse, which were thoroughly studied and can be used as a reference [5], but one should stress that the present results yield only first estimates, which need adaptation and revision when various measurement devices are considered. Quite often, there are different simultaneous carriers of intricacy with various velocities (*e.g.* the sound velocity, Fermi velocity or the velocity of light) and different intensities. But this is another matter.

### 9. Interpretation and conclusion

*What is the proper framework of interpretation*?

An interesting question arises when one tries to draw conclusions. It looks at first sight as a technical question regarding iteration, when one goes from the previous calculation of fluctuations during the time interval [*t* , *t* + δ*t*] to the next calculation for the interval [*t* + δ*t*, *t* + 2δ*t*]. The problem is then: From which set of values for ($p_1, p_2$) at time *t* + δ*t* should one start the iteration?

There are two options. The first one considers that the first step (the calculation in Section 8) yielded a large set of possible values for the quantities δ$p_1$, with a well-defined probability distribution. One could then start the next iteration from this distribution. The second option assumes that there exists on the contrary an actual couple of values ($p_1, p_2$) at every time, so that one should start iteration from these actual values as they are at time *t* + δ*t*. Both approaches lead to deep questions belonging to the interpretation of quantum mechanics. This is why they were deleted till coming near conclusions.

When one considers the first option, it will be shown soon that it leads to collapse through an accumulation of fluctuations, until a final value $p_1 = 1$ or $p_1 = 0$ is reached. The final outcome consists then in two distinct events where the couple ($p_1, p_2$) is equal either to (1, 0) or to (0, 1), with definite probabilities for the two eventualities. But it is far from obvious to spell out the meaning of these events. How can a unique outcome become real, and what does one mean then by "real"?

The second option digs deeper into interpretation since, although it hinges again on the question of reality, it does not ask this question at the end of a calculation but all along.
To envision the problem clearly, one must go back to another question, which remained tacit till now. It is concerned with the meaning of the density matrices with which one worked, namely: How are they defined?
A simple tentative answer consists in pushing the question far enough out of the measuring apparatus and the measured particle. One assumes then that the quantum state (as a density matrix) of a sufficiently large part of the universe makes sense. For the sake of consistency, this part of the universe should be large enough to include the alpha particle, the detector and an environment. The state $\rho_{AB}$ would be then defined by a partial trace over something much larger. According to Von Neumann, the quantities $(p_1, p_2)$ at time $t + \delta t$ would be average values of the observables $|1><1|$ and $|2><2|$ according to $\rho_{AB}(t + \delta t)$. They would be the quantities occurring in principle in the iterative algorithm.

But one cannot then avoid a next question, which is: How is defined the density matrix of the large embedding region? Willingly or not, sooner or later, one is pushed towards the troublesome question: What is the quantum state of the universe and should it be expressed by a universal wave function Ψ?

Many people believe that the existence of Ψ is sensible and many other people, perhaps more numerous, judge it metaphysical. There is however a less drastic position, which I adopt personally and says: *The idea of a wave function of the universe is not so much a metaphysical concept than a constraint on our present way of thinking. It is the only one allowing us to think of the universe when the only conceptual framework at our disposal is quantum theory and one must deal with reality, because the universe is the essence of reality for physicists.* This point of view recognizes the limits of our knowledge of the laws of nature and tries nevertheless to draw the most from what we know presently of these laws.

Assuming, or rather using the existence of Ψ as a way of speaking, does not imply that this wave function should divide itself into separate branches when a quantum measurement, or something analogous happens. Taking again as an example the experiment that is discussed here, the necessary assumptions could have been stated as saying that Ψ involved the existence of the alpha particle and, when tracing out everything else than the alpha particle in |Ψ><Ψ| one would have got the pure state (8.1).

As far as I can see, again personally, nothing in a quantum evolution of Ψ implies that the quantities $(p_1, p_2)$ should be invariant when the particle interacts with a non-isolated macroscopic part of the world. The only statement that one can make comes from decoherence theory [6] and recognizes that these quantities are invariant if the *past history* of the universe plays no part in the process, or if in other words one ignores predecoherence.

The existence of such a memory is the basic idea of intricacy, with its consequence in the existence and action of predecoherence extending the limits of action of decoherence. Intricacy is irreversible and the universe itself is a paradigm of irreversibility. Everything known about it shows that it keeps memory. If Ψ exists, it carries certainly a memory of itself. In that sense, the theory of intricacy deals with a little part of this memory in a little corner of the universe.

To say that again differently, one may notice that measurement theory is basically a problem in irreversibility, which starts from two separated systems *A* and *B* and asks what they become *later*. Even Schrödinger, when he discussed the concept of entanglement [2], had to use the words "before" and "after": Beforehand, *A* and *B* had not interacted, thereafter, they are entangled.

From this standpoint, the outcome of a quantum measurement, here and now, is an almost pointless event in the history of the universe, although it gives us much worry. It is contained in Ψ and predicted by the evolution of Ψ, with no chance for branching, because its influence travels at a finite velocity. Past states, including the collisions of some air molecules on the box of a Geiger counter; determined later states and some actual values of the quantities $(p_1, p_2)$ at every time, as well as implying a final collapse in which one of these squared amplitude vanishes. The present theory appears therefore as an attempt to find what happens when recognizing that Ψ is unattainable. The impossibility of knowing Ψ constrains us to rely on the mathematical paradigm of recognized ignorance, which is probability calculus

Another approach, rarely mentioned, expresses otherwise a necessary attitude of "learned ignorance". It relies on the recognition that nothing is known of the quantum laws of nature beyond a precision of $10^{-12}$, whereas much of the discussions on the quantum states of macroscopic systems –not even mentioning the universe– refer implicitly to a tremendously higher precision. One could then envision that there is a limit to our present mathematical expression of physics, so that the concept of wave functions would be intrinsically limited in precision. This idea, which could lead to many variants, would also be compatible with the kind of random collapse to which one arrives here.

Anyway, whatever interpretation and whatever option concerning the meaning of the algorithm one chooses, there seems to be no fundamental objection against assuming that the present approach to collapse is sensible. To conclude on this point, one will only say after many thinkers that quantum mechanics does not by itself predict Reality [19], but one should not abandon the expectation to find it consistent with Reality.

*Collapse*

The next question is more technical and asks how fluctuations imply collapse and consistency with a unique Reality at macroscopic scales. To deal with it, one can notice several points:

1. The fluctuations (8.12) are Brownian.
2. If one denotes by *A* the correlation coefficients in (8.12), they depend only upon the channel probabilities at time *t*, since the measures of intricacy $f_\beta$ entering in them depend only on time.
3. A derivative $\partial A/\partial p$ does not vanish when some *p* (either $p_1$ or $p_2$) vanishes: This means that a probability can vanish after a finite time.
4. The correlation A vanish when some *p* vanishes: This means that if a probability *p* vanishes at some time, it cannot revive later on because the fluctuations that could have revived it disappeared with it.

Altogether, these properties predict that there will be necessarily collapse in an individual measurement, according to a famous theorem by Pearle [20]. Collapse occurs when all the channel probabilities have vanished except one of them, which has become equal to 1. These events must behave randomly in a series of measurements, because the series of external collisions are never twice the same. Pearle's theorem, which is a long-shot version of Huygens' theorem on the gambler's ruin problem, shows thus an essential result, which is that the frequencies of final results in a long series of "identical" measurements must coincide with the predictions of the Born probability rule.

*Non-separability*

Another problem of consistency arises when the measuring apparatus *B* is not unique, for instance when the alpha particle can be detected elsewhere in state $|2>$ by another apparatus. This is a fundamental question since it asks why the results of two measurements, both local though distant, should agree with the correlations in the initial quantum state of the measured system. This is of course the problem, or the character of non-separability in quantum mechanics. The answer is easily obtained through an adaptation of Section 8 taking both measuring apparatuses into account, as one did when using cells *β*. It relies on Pearle's theorem and was already published elsewhere [18].

*Conclusion*

The collapse mechanism, which has been introduced here, is certainly new. It should be compared however with Zurek's previous proposal of Quantum Darwinism, according to which collapse would occur in the environment and not in the measuring device [21]. One may notice that measures of intricacy, as they were introduced here, could be helpful also in the framework of Quantum Darwinism and provide a more versatile tool than relative measures of algorithmic information, which were used by Zurek.

As far as the present theory is concerned, one cannot presently exclude the possibility of some unspotted errors in its construction or misinterpretations of its meaning, but it seems nevertheless that there could be something valuable in it, although some of its remaining problems exceed by far the abilities or possibilities of the present author.